\documentclass[a4paper,twoside]{article}
\usepackage[pdftex]{graphicx}
\usepackage{epsfig}
\usepackage{subcaption}
\usepackage{calc}
\usepackage{amssymb}
\usepackage{amstext}
\usepackage{amsmath}
\usepackage{amsthm}
\usepackage{multicol}
\usepackage{pslatex}
\usepackage{apalike}
\usepackage[bottom]{footmisc}
\usepackage{SCITEPRESS}
\usepackage{nccmath}
\usepackage{tikz}
\usepackage{here}
\usepackage{mathrsfs}
\usepackage{mathtools}

\usetikzlibrary{arrows.meta}

\theoremstyle{plain}
\newtheorem{df}{Definition}

\newtheorem{thm}{Theorem}
\newtheorem*{exa}{Example}

\newtheorem{lemma}{Lemma}

\begin{document}

\title{Logic of Awareness in Agent's Reasoning}

\author{
  \authorname{Yudai Kubono\orcidAuthor{0000-0003-2617-8870}, 
  Teeradaj Racharak\orcidAuthor{0000-0002-8823-2361} 
  and Satoshi Tojo}
  \affiliation{School of Information Science,
   Japan Advanced Institute of Science and Technology, Ishikawa, Japan}
  \email{\{k-yudai, racharak, tojo\}@jaist.ac.jp}}


\abstract{
  The aim of this study is to formally express 
  awareness for modeling practical agent communication. 
  The notion of awareness has been proposed as a set 
  of propositions for each agent, 
  to which he/she pays attention, and has contributed to 
  avoiding \textit{logical omniscience}. 
  However, when an agent guesses another agent's 
  knowledge states, what matters are not propositions 
  but are accessible possible worlds. 
  Therefore, we introduce a partition of possible 
  worlds connected to awareness, 
  that is an equivalence relation, 
  to denote \textit{indistinguishable} worlds.
  Our logic is called Awareness Logic with Partition ($\mathcal{ALP}$). 
  In this paper, we first show a running example to 
  illustrate a practical social game. 
  Thereafter, we introduce syntax and Kripke semantics 
  of the logic and prove its completeness. 
  Finally, we outline an idea to incorporate 
  some epistemic actions with 
  dynamic operators that change the state of awareness.}

\onecolumn \maketitle \normalsize \setcounter{footnote}{0} 

\section{\uppercase{Introduction}}
\label{sec:introduction}
In a society of rational agents, 
communication among them can be defined 
by means of message's exchanges in which 
each message is represented by a logical formula. 
In this context, a recipient agent 
may change or revise his/her belief according to the received 
message to maintain the logical consistency of knowledge.

First, we denote a unit of knowledge by $\varphi,\psi,\cdots$ 
and write $K_a\varphi$ for `agent $a$ knows $\varphi$,' 
or we may write anonymously $K\varphi$. 
In such a formalization, \textit{logical omniscience} matters; 
in ordinary logic, we employ \textit{Modus Ponens} 
(MP)\footnote{From $\varphi$ and $\varphi\to\psi$, we conclude $\psi$.}
for logical inference, and when one knows $\varphi$ and 
$\varphi \to \psi$, i.e., $K\varphi$ and $K(\varphi\to\psi)$, respectively,
$K\psi$ would necessarily be inferred in his/her knowledge 
if we adopt the axiom 
K.\footnote{K: $K(\varphi\to\psi) \to (K\varphi \to K\psi)$.} 
However, such exhaustive reasoning is unrealistic for human model.
The logic of an agent's knowledge/belief 
is called \textit{epistemic logic}, and
its semantics is given by a Kriple model that consists of
a set of possible worlds where each world has 
different valuation for propositions,
to which each agent may or may not be accessible.
When an agent can access two worlds of different valuations,
e.g., each of which includes $\varphi$ or $\neg\varphi$, 
he/she does not know whether $\varphi$ is true or not, 
that is, $\neg K \varphi$. 
On the contrary, when an agent can find $\varphi$ 
in all his/her accessible worlds, $K \varphi$ holds.
To avoid logical omniscience, 
we need to restrict propositions 
to be employed in reasoning, apart from those not to be employed.

\begin{figure}[b]
  \centering
  \includegraphics[width=7.5cm]{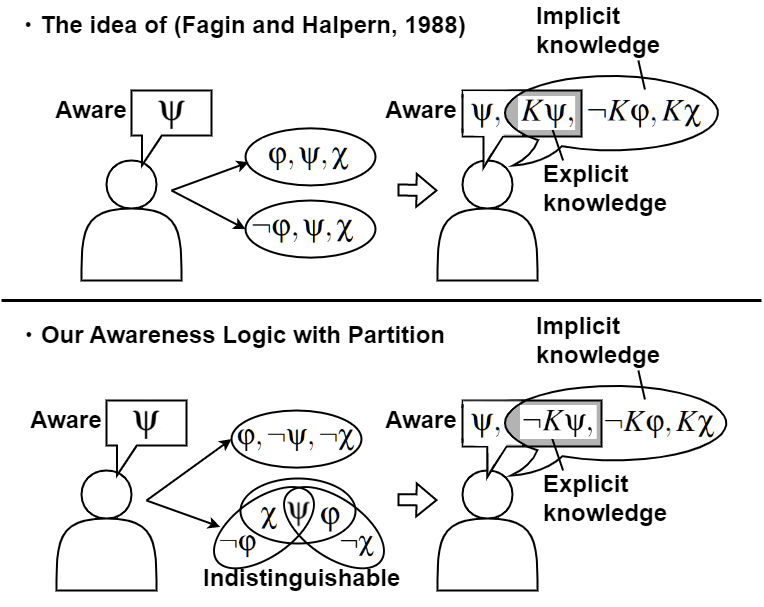}
  \caption{A comparison on the intuitions of 
  the previous study and this paper.}
\end{figure}

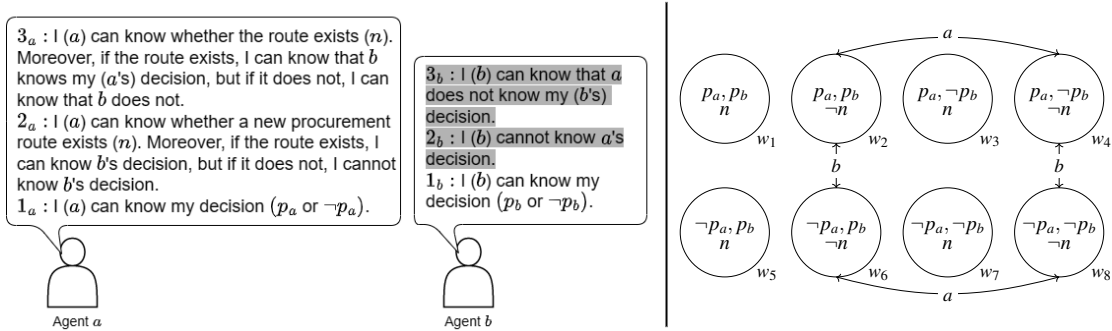
\begin{figure*}[h]
  \begin{tabular}{cc}
  \begin{minipage}{0.58\hsize}
    \includegraphics[width=9.35cm]{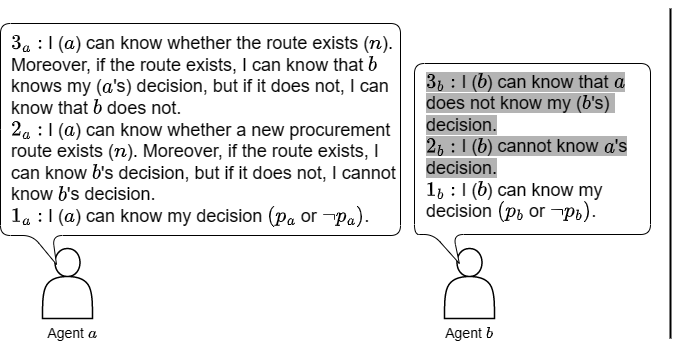}
    \end{minipage}
    \begin{minipage}{0.42\hsize}
    \centering
    \scalebox{0.63}[0.63]{
    \begin{tikzpicture}
    \draw(0,0)circle(1);
    \draw(0.95,-0.95)node{{\large $w_1$}};
    \draw(2.5,0)circle(1);
    \draw(3.45,-0.95)node{{\large $w_2$}};
    \draw(5,0)circle(1);
    \draw(5.95,-0.95)node{{\large $w_3$}};
    \draw(7.5,0)circle(1);
    \draw(8.45,-0.95)node{{\large $w_4$}};
    \draw(0,-3)circle(1);
    \draw(0.95,-3.95)node{{\large $w_5$}};
    \draw(2.5,-3)circle(1);
    \draw(3.45,-3.95)node{{\large $w_6$}};
    \draw(5,-3)circle(1);
    \draw(5.95,-3.95)node{{\large $w_7$}};
    \draw(7.5,-3)circle(1);
    \draw(8.45,-3.95)node{{\large $w_8$}};
    \draw(0,0.1)node{{\Large $p_a,p_b$}};
    \draw(0,-0.05)node[below]{{\Large $n$}};
    \draw(2.5,0.1)node{{\Large $p_a,p_b$}};
    \draw(2.5,-0.05)node[below]{{\Large $\neg n$}};
    \draw(5,0.1)node{{\Large $p_a,\neg p_b$}};
    \draw(5,-0.05)node[below]{{\Large $n$}};
    \draw(7.5,0.1)node{{\Large $p_a,\neg p_b$}};
    \draw(7.5,-0.05)node[below]{{\Large $\neg n$}};
    \draw(0,-2.9)node{{\Large $\neg p_a, p_b$}};
    \draw(0,-3.05)node[below]{{\Large $n$}};
    \draw(2.5,-2.9)node{{\Large $\neg p_a,p_b$}};
    \draw(2.5,-3.05)node[below]{{\Large $\neg n$}};
    \draw(5,-2.9)node{{\Large $\neg p_a,\neg p_b$}};
    \draw(5,-3.05)node[below]{{\Large $n$}};
    \draw(7.5,-2.9)node{{\Large $\neg p_a,\neg p_b$}};
    \draw(7.5,-3.05)node[below]{{\Large $\neg n$}};
    \draw[<->](2.5,1)to [out=15, in=165] (7.5,1);
    \draw(5,1.45)node[fill=white]{{\large $a$}};
    \draw[<->](2.5,-4)to [out=-15, in=-165] (7.5,-4);
    \draw(5,-4.45)node[fill=white]{{\large $a$}};
    \draw[<->](2.5,-1)--(2.5,-2);
    \draw(2.5,-1.5)node[fill=white]{{\large $b$}};
    \draw[<->](7.5,-1)--(7.5,-2);
    \draw(7.5,-1.5)node[fill=white]{{\large $b$}};
    \end{tikzpicture}\par}
  \end{minipage}
  \end{tabular}
\caption{The left side: each agent's knowledge and their reasoning, 
which affect their own decisions. 
The right side: the Kripke model from $a$'s viewpoint.}
\end{figure*}

\cite{fagin1988belief} proposed 
components that represent agents' state of 
awareness called an \textit{awareness set} and 
incorporated it into epistemic logic. 
This logic distinguishes the knowledge 
that the agents cannot use for their reasoning, 
called \textit{implicit knowledge}, from 
that they can, called \textit{explicit knowledge}. 
The former, implicit knowledge, represents unaware information. 
The idea of \cite{fagin1988belief} is 
to classify knowledge into implicit or explicit knowledge 
according to whether an agent is aware of the proposition (Figure 1). 
It is a simple, intuitive definition 
and the main approach in logic of awareness. 

However, we argue that awareness 
should also affect the distinction of possible worlds 
in addition to propositions. 
In the previous study, 
awareness only concerns the propositions. 
However, when an agent is unaware of a certain proposition,  
he/she must not also be aware of 
the distinction of two possible worlds.
In figure 1, the agent accesses only the possible worlds 
where $\chi$ is true despite unawareness of $\chi$.
This indistinction plays an important role when 
agents make inferences about other agents' knowledge states, 
as shown in our illustrative example in Section 2. 
Therefore, in this paper, we propose a framework to 
mention that two possible worlds are indistinguishable 
from the viewpoint of an agent.
Our logic can allow us to handle reasoning correctly 
about what knowledge other agents have 
and also enables to formalize 
practical agent communications. 

Besides, in game theory, players make their own decisions
by guessing other players' reasoning. 
It is based on specific decision criteria, such as 
the best strategy to a dominant strategy. 
Whether player $b$ is aware of actions 
that player $a$ can take affects $b$'s strategy 
to find an equilibrium.
In this sense, our logic is supposed to be useful 
in its application to game theory. 

The paper is structured as follows.
In Section 2, we introduce an example about 
inferences among multi-agents that shows the necessity 
for introducing our logic.
In Section 3, we introduce \textit{Awareness Logic with Partition} 
($\mathcal{ALP}$), which is based on 
Awareness Logic \cite{fagin1988belief}. 
Its semantics was given in the Kripke-style. 
We add a new equivalence relation, which is connected 
to the states of an agent's awareness from another agent's 
viewpoint, to the standard Kripke model.
Besides, we show how our logic 
works using the example presented in Section 2.
In Section 4, we give a proof system \textbf{ALP} of our logic 
$\mathcal{ALP}$ in Hilbert-style. 
As for proving the completeness theorem, we use techniques of 
logic of the modality for transitive closure \cite{van2007dynamic}. 
Section 5 discusses two epistemic actions: 
\textit{becoming aware of} and \textit{becoming unaware of}, 
and gives an extension of $\mathcal{ALP}$. 
In Section 6, we introduce some related work. 
Section 7 concludes.

\section{\uppercase{Example: Convenience store's expansion}}
This section gives an example at convenience stores. 
It describes a situation where agents have 
different states of awareness. 
\begin{exa}
  Let agent $a$ be the owner of the convenience 
  store $A$ and agent $b$ be the owner of 
  the convenience store $B$ 
  considering to open his/her own new store. 
  The cost of products has risen due to poor harvests, 
  and a reckless expansion leads to a significant loss. 
  Agent $a$ is aware of a new procurement route that 
  allows the owners to purchase products 
  at half the current price. 
  Moreover, $a$ is unaware that $b$ 
  is unaware of the existence of the new route. 
\end{exa}
In this example, if owner $a$ can know $b$'s decision, 
that decision can be a helpful factor in $a$'s decision.
For example, if owner $b$ decides to open a new store, 
owner $a$ is also likely to decide to expand 
a new store because otherwise, it may be disadvantageous.

We denote $p_a$ and $p_b$ for the propositions that $a$ and $b$ 
expand their stores, respectively, and $n$ 
for the proposition that there is a new procurement route.
As for the agents' knowledge, 
$a$ can know $p_a$ and $p_b$ in the case that $n$ is true.  
On the other hand, 
$b$ can know only $p_b$ in any case because he/she does not 
have a clue about $a$'s knowledge state. 
Then, we give the Kripke model in Figure 2 
to see how each agent guesses the opponent's knowledge. 
Note that we omit reflexivity on accessibility relations 
from a figure for visibility.
Basically, each agent does not know the decision of 
the opponent, i.e., $b$'s decision for $a$ and 
$a$'s decision for $b$ are unknown to each other. 
However, in possible worlds where $n$ holds, 
$a$ can know $p_b$, and $b$ can know $p_a$. 
Thus, there is no accessibility relation between 
possible worlds such that it has a different valuation for 
each $p_b$ and $p_a$ and $n$ is true. 
As for states of awareness, 
$a$ is aware of $p_a,p_b$, and $n$. 
However, $b$ is unaware of $n$. 
We can summarize it as awareness sets 
$\mathscr{A}_a=\{p_a,p_b,n\}$ for $a$ and 
$\mathscr{A}_b=\{p_a,p_b\}$ for $b$. 

Note that $2_b$ and $3_b$ in Figure 2 
cannot be correctly represented by the existing method.
Let $K_a$ and $K_b$ be operators 
expressing each explicit knowledge. 
We consider the truth value of $K_b K_a p_b$ at $w_1$, 
which means we assume $w_1$ as the actual world. 
In the method of \cite{fagin1988belief}, 
a $K_i$ operator for each agent $i$ is defined to be satisfied 
when the proposition is included in an awareness set and holds 
in all the accessible worlds (Figure 1). 
Thus, $K_b K_a p_b$ at $w_1$ implies that 
$p_b$ is contained in $b$'s awareness set, 
and $K_a p_b$ holds at $w_1$ 
that is accessible from $w_1$ on the edge labeled $b$. 
Also, $K_a p_b$ implies that 
$p_b$ is contained in $a$'s awareness set, 
and $p_b$ holds at $w_1$ that is accessible 
from $w_1$ on the edge labeled $a$. 
Since $p_b$ holds in $w_1$, the formula holds at $w_1$. 

However, $K_b K_a p_b$ contradicts $3_b$ in Figure 2, and 
our intuition considers that this formula should not hold at $w_1$. 
It is because $b$ is unaware of $n$ and cannot think that 
$a$ has a way of knowing $p_b$ at all. 
This unawareness means that some worlds, 
such as $w_1$ and $w_2$, 
are indistinguishable from $b$'s viewpoint. 
Therefore, the Kripke model from $b$'s viewpoint 
should be the form of Figure 3, in which $K_b K_a p_b$ 
must be evaluated by whether $p_b$ holds 
in all possible worlds that are $w_1',w_2',w_3'$ and $w_4'$. 
Thus, $K_b K_a p_b$ does not hold. 
Figure 2 is a Kripke model from $a$'s viewpoint 
who is aware of all atomic propositions discussed, 
which is different from that of $b$, 
because atomic propositions that $a$ is aware of 
are different from of those $b$ is aware.
Besides, a formula $K_b p_a$ also holds at $w_1$ and 
contradicts $2_b$ for the same reason. 

\begin{figure}
\centering
\scalebox{0.65}[0.65]{
\begin{tikzpicture}
\draw(0,0)circle(1);
\draw(0.95,-0.95)node{{\large $w_1'$}};
\draw(3,0)circle(1);
\draw(3.95,-0.95)node{{\large $w_2'$}};
\draw(0,-3)circle(1);
\draw(0.95,-3.95)node{{\large $w_3'$}};
\draw(3,-3)circle(1);
\draw(3.95,-3.95)node{{\large $w_4'$}};
\draw(0,0)node{{\Large $p_a,p_b$}};
\draw(3,0)node{{\Large $ p_a,\neg p_b$}};
\draw(0,-3)node{{\Large $\neg p_a,p_b$}};
\draw(3,-3)node{{\Large $\neg p_a,\neg p_b$}};
\draw[<->](1,0)--(2,0);
\draw(1.5,0)node[fill=white]{{\large $a$}};
\draw[<->](1,-3)--(2,-3);
\draw(1.5,-3)node[fill=white]{{\large $a$}};
\draw[<->](0,-1)--(0,-2);
\draw(0,-1.5)node[fill=white]{{\large $b$}};
\draw[<->](3,-1)--(3,-2);
\draw(3,-1.5)node[fill=white]{{\large $b$}};
\end{tikzpicture}\par}
\caption{The Kripke model from $b$'s viewpoint.}
\end{figure}
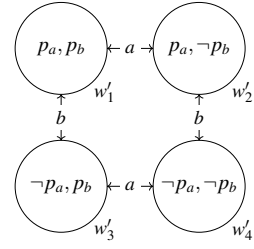

As we have seen, the existing method 
cannot treat some information 
used in the decision-making.
In order to correctly represent 
inference of agents, 
it is necessary not only to classify knowledge 
using states of awareness about propositions, 
but also to consider the distinction of 
possible worlds corresponding to agents' viewpoints.

Models that can distinguish between \textit{awareness of} and 
\textit{awareness that} has been proposed in the field of philosophy. 
The models allow us to represent 
a more accurate description of situations 
\cite{grossi2015syntactic,fernandez2021awareness}. 
The former is awareness 
in the sense of being able to refer to the information. 
The latter is awareness in the sense of 
acknowledging that the information is true 
through reasoning or observation. 
Although both concepts are similar, they have different properties. 
Previously proposed logic 
defines explicit knowledge by combining these two concepts. 
In this paper, we focus on `awareness of.' 
This is because the concept is more relevant to the example, 
and we do not need to consider the other one. 

\section{\uppercase{Awareness Logic with Partition}}
\subsection{Language}
\noindent
First of all, we define the syntax of $\mathcal{ALP}$.
Let $\mathcal{P}$ be a countable set of atomic propositions and
$\mathcal{G}$ be a finite set of agents. 
The language $\mathcal{L}_{\mathcal{P}}$ 
is the set of formulas 
generated by the following grammar:
\begin{align*}
\mathcal{L}_{\mathcal{P}} \ni\varphi
::= \ &p \mid \neg\varphi \mid \varphi\wedge\varphi \mid 
A^i_j\varphi \mid\\[-2pt]
    &L_j\varphi\mid [\equiv]^i_j\varphi
\mid C^i_j\varphi\mid K^i_j\varphi,  
\end{align*}
where $p \in\mathcal{P}$
and $i,j \in \mathcal{G}$. 
Other logical connectives $\vee$, $\to$, 
and $\leftrightarrow$ are defined in the usual manner.

We call $A^i_j$, $L_j$, and $K^i_j$ 
as an awareness operator,  
an implicit knowledge operator, 
and an explicit knowledge operator, respectively. 
Notationally, 
\begin{itemize}
  \item $A^i_j\varphi$ means $\varphi$ is information that 
  $j$ is aware of from $i$'s viewpoint.
  \item $L_j\varphi$ means that $\varphi$ is 
  $j$'s implicit knowledge.
  \item $K^i_j\varphi$ means that $\varphi$ is $j$'s 
  explicit knowledge from agent $i$'s viewpoint.
\end{itemize}
$[\equiv]^i_j$ and $C^i_j$ are special operators 
introduced to define explicit knowledge and 
used as the basis for proofs studied in this paper.
The former operator means that 
the information is true at $j$'s 
state of awareness from $i$'s viewpoint. 
The latter means that $\varphi$ is a kind of $j$'s 
implicit knowledge from agent $i$'s viewpoint. 
Note that implicit knowledge referred to by $C^i_j\varphi$ 
is stronger than that referred to by $L_j\varphi$. 
It might be interesting to explore the relationship of these two operators; 
however, it is outside our scope and remains as our future task. 

\subsection{Semantics}
\noindent
Now, we move on to the semantics of $\mathcal{ALP}$. 
\begin{df}
   An \textit{epistemic model with awareness} $M$
   is a tuple $\langle W, \{R_i\}_{i\in\mathcal{G}}, V,
   \{\mathscr{A}^i_j\}_{i,j\in\mathcal{G}}, 
   \{\equiv^i_j\}_{i,j\in\mathcal{G}}
   \rangle$ consists of a domain $W$, 
   a set of accessibility relations $R_i$, 
   a valuation function $V$, 
   a set of awareness sets $\mathscr{A}^i_j$, 
   and a set of relations $\equiv^i_j$,   
   where:
   \begin{itemize}
    \item $W$ \text{ is a non-empty set of possible worlds}; 
    \item $R_i \subseteq W\times W$ 
    \text{ is an equivalence relation on }W;
    \item $V : \mathcal{P} \to 2^{W}$;
    \item $\mathscr{A}^i_j$ \text{ is a non-empty 
    set of atomic propositions }
    \item[] \text{satisfying that }$\mathscr{A}^i_j \subseteq\mathscr{A}^i_i$;
    \item $(w,v) \in \ \equiv^i_j
        \textit{\  iff  \ } 
        (w \in V(p) \textit{\ iff \ }
        v \in V(p) \ \ \ \ \ \ \ \ \ \ \ \ \ \ $
        $\text{ for every } p \in \mathscr{A}^i_j)$.   
   \end{itemize}
 \end{df}
The pair $(M,w)$ with $M$ and $w\in W$ 
in it is called a pointed model. 
We can say that 
$i$'s viewpoint is formally an epistemic 
model with awareness where the superscript 
index is restricted to $i$, 
that is 
$\langle W, \{R_j\}_{j\in\mathcal{G}}, V, 
\{\mathscr{A}^i_j\}_{j\in\mathcal{G}}, 
\{\equiv^i_j\}_{j\in\mathcal{G}} \rangle$. 
We call the restricted model 
$i$'s epistemic model with awareness. 

The condition $\mathscr{A}^i_j \subseteq\mathscr{A}^i_i$
means that atomic propositions of which 
$j$ is aware from $i$'s viewpoint do not 
contain a proposition of which $i$ himself/herself is unaware. 
We call $\equiv^i_j$ \textit{indistinguishable relations} for $j$ 
from $i$'s viewpoint. 
An indistinguishable relation 
$\equiv^i_j$ is a relation between possible 
worlds with a different valuation
for atomic propositions that $j$ is unaware of 
from $i$'s viewpoint. 
This represents that, from $i$'s viewpoint, 
$j$ cannot distinguish such possible worlds. 
By partitioning $W$ using an indistinguishable relation, 
we can formalize knowledge according to the propositions 
of which the agent is aware. 
Possible worlds that are indistinguishable because of being unaware 
are collapsed with an equivalence class.

Note that 
there are local and global definitions of an awareness set. 
The former defines $\mathscr{A}_i$ as a function 
that takes a possible world as an argument and 
changes elements of an awareness set for each possible world.
The latter defines an awareness set as the same in 
all possible worlds.
Generally, a state of awareness is 
fixed within an agent's scope, which 
is a set of the agent's accessible possible worlds. 
Thus, a global definition is used in the logic that 
does not consider the outside of a specific agent's scope, 
such as a single-agent case. 
On the other hand, a local definition 
can represent a state of awareness 
in possible worlds outside the agent's scope. 
It is possible to express the possibility 
that there is a difference between the state of an agent's awareness 
in his/her scope and that in other agents' scope. 

This logic adopts the global one, 
because even with the global definition, 
it is possible to express the possibility that 
the state of an agent's awareness from his/her viewpoint 
is different from that from other agents' viewpoints, 
which is an advantage of the local definition.
It follows easily from the definition that every 
state of awareness is uniquely set for each agent. 

We move on to the satisfaction relation.
At first, we introduce some notations for the definition: 
$At(\varphi)$ is denoted as
the set of atomic propositions that appear in $\varphi$; 
$R_j \ \circ\equiv^i_j$ is denoted as a sequential composition 
of $\equiv^i_j$ and $R_j$; 
$R^+$ is denoted as the transitive closure of 
$R$. This $R^+$ is the smallest set such that $R\subseteq R^+$, 
and for all $x, y, z$, 
if $(x,y)\in R^+$ and $(y,z)\in R^+$, then $(x,z)\in R^+$. 
  \begin{df}
    For any epistemic models with awareness
    $M$ and possible worlds $w \in W$,  
    the satisfaction relation $\vDash$ is given as follows: 
    \begin{align*}
     M,w \vDash p &\textit{\  iff  \ } w \in V(p) 
      ; \\[-2pt]
     M,w \vDash \neg \varphi &\textit{\  iff  \ } M,w 
     \nvDash\varphi;\\[-2pt]
     M,w \vDash \varphi\wedge\psi &\textit{\  iff  \ } 
     M,w\vDash\varphi \text{, and } M,w\vDash\psi ;\\[-2pt]
     M,w \vDash A^i_j 
     \varphi &\textit{\  iff  \ } At(\varphi) 
     \subseteq \mathscr{A}^i_j;\\[-2pt]
    M,w \vDash L_j\varphi &\textit{\  iff  \ } 
    M,v\vDash \varphi \text{ for all } v \\[-4pt]
    &\hspace{2.4ex}\text{ such that }
    (w,v)\in R_j;\\[-2pt]
    M,w\vDash [\equiv]^i_j\varphi &\textit{\  iff  \ } 
    M,v\vDash\varphi \text{ for all } v \\[-6pt]
    &\hspace{2.4ex}\text{ such that }
    (w,v)\in \ \equiv^i_j;\\[-2pt]
    M,w\vDash C^i_j\varphi &\textit{\  iff  \ } 
    M,v\vDash \varphi \text{ for all } v\\[-6pt] 
    &\hspace{2.4ex}
    \text{ such that } (w,v) \in (R_j \ \circ\equiv^i_j)^+;\\[-2pt]
     M,w \vDash K^i_j \varphi &\textit{\  iff  \ } 
     M,w \vDash A^i_j\varphi  \text{, and }
     M,w\vDash C^i_j\varphi.
    \end{align*}
    \end{df}

From Definition 1, it spells out 
that if both indistinguishable relations 
$\equiv^i_j$ and accessibility relations $R_j$ is equivalent, 
then $( R_j \ \circ\equiv^i_j)^+$ is equivalent. 
Since both relations are equivalence relations, 
the reverse direction on the composition is also reachable, 
although it consumes a few extra steps.
Thus, $( R_j \ \circ\equiv^i_j)^+$ gives a new partition 
of possible worlds. 
From the definitions, we can also find that 
$[\equiv]^i_j L_j\varphi$ corresponds to $R_j \ \circ\equiv^i_j$. 
However, this relation is not equivalent, 
unlike its transitive closure. 
    
Next, we define the validity in the usual way.  
  \begin{df}
    A formula $\varphi$ is valid at $M$, 
    if $\varphi$ holds at every pointed model $M,w$ in $M$, 
    which is denoted by $M\vDash\varphi$. 
    A formula $\varphi$ is valid
    if $\varphi$ holds at every pointed model $M,w$, 
    which is denoted by $\vDash \varphi$. 
  \end{df}

\subsection{FORMALIZATION OF THE EXAMPLE}
We formalize the running example using our logic 
and consider the truth values of $K_b K_a p_b$ again. 
Agent $b$ is unaware of $n$. 
Then there are indistinguishable relations 
between possible worlds where $n$ holds or not, 
such as $w_1$ and $w_2$. 
Formally, we write $\equiv^a_a$ and $\equiv^a_b$ 
as $\emptyset$. 
We also formalize $\equiv^b_a$ and $\equiv^b_b$ 
as $\{(w_1,w_2),(w_3,w_4),
(w_5,w_6),(w_7,w_8),(w_2,w_1),$\\$(w_4,w_3),
(w_6,w_5),(w_8,w_7),(w_1,w_1),\cdots,(w_8,w_8)\}$.
In our logic, $K_b K_a p_b$ is rewritten as the form 
$K^b_b K^b_a p_b$ by introducing agents' viewpoint. 
This formula is evaluated by whether $p_b$ holds 
in all the reachable worlds on 
$\equiv^b_b$, $R_b$, $\equiv^b_a$, and $R_a$. 
As seen in Figure 4, since it is false in some worlds, 
the formula does not hold at $w_1$. 
On the other hand, $K^b_b \neg K^b_a p_b$, 
which is consistent with $3_b$ in Figure 2, is true.

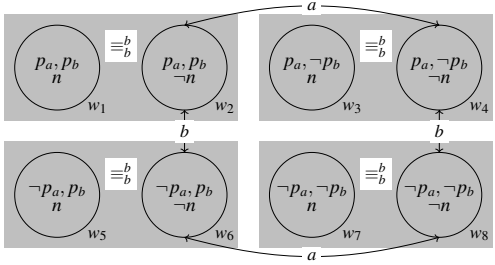
\begin{figure}
\centering
\scalebox{0.6}[0.6]{
\begin{tikzpicture}
\fill[lightgray](-1.25,-1.25)rectangle(4.25,1.25);
\fill[lightgray](4.75,-1.25)rectangle(10.25,1.25);
\fill[lightgray](-1.25,-1.75)rectangle(4.25,-4.25);
\fill[lightgray](4.75,-1.75)rectangle(10.25,-4.25);
\draw(1.5,0.5)node[fill=white]{{\large $\equiv^b_b$}};
\draw(7.5,0.5)node[fill=white]{{\large $\equiv^b_b$}};
\draw(1.5,-2.5)node[fill=white]{{\large $\equiv^b_b$}};
\draw(7.5,-2.5)node[fill=white]{{\large $\equiv^b_b$}};
\draw(0,0)circle(1);
\draw(0.95,-0.95)node{{\large $w_1$}};
\draw(3,0)circle(1);
\draw(3.95,-0.95)node{{\large $w_2$}};
\draw(6,0)circle(1);
\draw(6.95,-0.95)node{{\large $w_3$}};
\draw(9,0)circle(1);
\draw(9.95,-0.95)node{{\large $w_4$}};
\draw(0,-3)circle(1);
\draw(0.95,-3.95)node{{\large $w_5$}};
\draw(3,-3)circle(1);
\draw(3.95,-3.95)node{{\large $w_6$}};
\draw(6,-3)circle(1);
\draw(6.95,-3.95)node{{\large $w_7$}};
\draw(9,-3)circle(1);
\draw(9.95,-3.95)node{{\large $w_8$}};
\draw(0,0.1)node{{\Large $p_a,p_b$}};
\draw(0,-0.05)node[below]{{\Large $n$}};
\draw(3,0.1)node{{\Large $p_a,p_b$}};
\draw(3,-0.05)node[below]{{\Large $\neg n$}};
\draw(6,0.1)node{{\Large $p_a,\neg p_b$}};
\draw(6,-0.05)node[below]{{\Large $n$}};
\draw(9,0.1)node{{\Large $p_a,\neg p_b$}};
\draw(9,-0.05)node[below]{{\Large $\neg n$}};
\draw(0,-2.9)node{{\Large $\neg p_a, p_b$}};
\draw(0,-3.05)node[below]{{\Large $n$}};
\draw(3,-2.9)node{{\Large $\neg p_a,p_b$}};
\draw(3,-3.05)node[below]{{\Large $\neg n$}};
\draw(6,-2.9)node{{\Large $\neg p_a,\neg p_b$}};
\draw(6,-3.05)node[below]{{\Large $n$}};
\draw(9,-2.9)node{{\Large $\neg p_a,\neg p_b$}};
\draw(9,-3.05)node[below]{{\Large $\neg n$}};
\draw[<->](3,1)to [out=15, in=165] (9,1);
\draw(6,1.45)node[fill=white]{{\large $a$}};
\draw[<->](3,-4)to [out=-15, in=-165] (9,-4);
\draw(6,-4.45)node[fill=white]{{\large $a$}};
\draw[<->](3,-1)--(3,-2);
\draw(3,-1.5)node[fill=white]{{\large $b$}};
\draw[<->](9,-1)--(9,-2);
\draw(9,-1.5)node[fill=white]{{\large $b$}};
\end{tikzpicture}\par}\caption{The Kripke model depicted by $\mathcal{ALP}$.}
\end{figure}

As for other formulas, such as $\neg K^b_b p_a$, 
which is consistent with $2_b$, 
and $K^b_a K^b_b \neg K^b_a p_b$, 
these are also consistent with our intuition. 
The latter means that from $b$'s viewpoint, 
$a$ knows that $b$ knows that $a$ does not know $p_b$.
A formula $K^a_a K^a_b \neg K^a_a p_b$, 
which has the same meaning from $a$'s viewpoint, 
does not hold at $w_1$.
Moreover, we can consider a situation where 
$a$ is aware that $b$ is unaware of $n$ and 
express the situation by formalizing 
$\mathscr{A}^a_b$ as $\{p_a,p_b\}$. 
$K^a_a K^a_b \neg K^a_a p_b$ is true at $w_1$, 
which is $a$ knows $b$'s incorrect knowledge. 

In Figure 4, the equivalence classes of 
the indistinguishable relation $\equiv^b_b$ 
are represented by the light gray background. 
By interpreting the equivalence class 
as one possible world from $b$'s viewpoint, 
Figure 4 represents the same graph 
as the Kripke model from $b$'s viewpoint in Figure 3 
in terms of possible worlds and accessibility relations. 
Thus, we can say that 
our logic represents the distinction of possible worlds 
according to states of awareness for each agent. 
\section{HILBERT-SYSTEM FOR $\mathcal{ALP}$}
We now move on to the proof theory for $\mathcal{ALP}$. 
The Hilbert-system of our logic is given in Table 1. 
AN, AC, AA, AL, $\mathrm{A[\equiv]}$, $\mathrm{ACM}$, and AK 
mean that if an agent is aware of atomic propositions, 
he/she is aware of more complex formulas produced by 
the atomic propositions 
and correspond to the meaning of `awareness of.'
For $\mathrm{K_L, T_L, 5_L, K_{[\equiv]}, T_{[\equiv]}},$ 
and $\mathrm{5_{[\equiv]}}$, 
we adopt K,T, and 5 axioms in modal logic. 
$\mathrm{5_L}$ called negative introspection 
in epistemic logic means that an agent always knows 
what he/she does not know. 
This axiom also characterizes logical omniscience. 
In our logic, as with most logics of awareness, 
this formula does not hold for $K^i_j$ operators.
Instead, $\neg K^i_j\varphi\wedge A^i_j\neg K^i_j\varphi
\to K^i_j \neg K^i_j\varphi$ is valid. 
$\mathrm{K_C}, \mathrm{IND}$, and $\mathrm{MIX}$ are based 
on axioms of logic with common knowledge \cite{fagin1995reasoning}, 
because the idea of transitive closure is the same as that one.
$\mathrm{KAC}$ corresponds to the 
definition of satisfaction relation of $K^i_j$. 
It means that explicit knowledge is the things 
that meet implicit knowledge referred 
to by $C^i_j$ and aware propositions.

\begin{df}
A system $\mathbf{ALP}$ is a set of formulas 
that contains the axioms in Table 1 
and is closed under inference rules in it. 
We write $\vdash\varphi$ if $\varphi\in$ $\mathbf{ALP}$. 
Let $\Gamma$ be a set of formulas in $\mathbf{ALP}$ and 
$\bigwedge\Gamma$ be an abbreviation of 
$\bigwedge_{\varphi\in \Gamma}\varphi$. 
If there is a finite subset $\Gamma'$ of $\Gamma$ such that 
$\vdash\bigwedge\Gamma'\to \varphi$, 
we write $\Gamma\vdash\varphi$ 
and call $\varphi$ derivation from $\Gamma$.
\end{df}

\begin{table}[h]
  \centering
  \caption{Axiom schemas and inference rules of \textbf{ALP}.}
   \begin{tabular}{|l|l|}
    \hline
    \multicolumn{2}{|c|}{Axioms}\\\hline
    TAUT & The set of propositional tautologies\\
    AN & $\vdash A^i_j\varphi \leftrightarrow A^i_j\neg\varphi$\\
    AC & $\vdash A^i_j(\varphi\wedge\psi) \leftrightarrow 
    A^i_j\varphi \wedge A^i_j\psi$\\
    AA & $\vdash A^i_j\varphi \leftrightarrow A^i_j A^k_l\varphi$\\
    $\mathrm{A[\equiv]}$ & 
    $\vdash A^i_j\varphi \leftrightarrow A^i_j [\equiv]^k_l\varphi$\\
    $\mathrm{ACM}$ & 
    $\vdash A^i_j\varphi \leftrightarrow A^i_j C^k_l\varphi$\\
    AL & $\vdash A^i_j \varphi\leftrightarrow A^i_j L_k \varphi$\\
    AK & $\vdash A^i_j \varphi \leftrightarrow A^i_j K^k_l \varphi$\\
    $\mathrm{AN[\equiv]}$ & $\vdash A^i_j p\wedge p
     \to [\equiv]^i_j p$\\
     $\mathrm{AI}$ &$\vdash A^i_j \varphi\to A^i_i \varphi$\\
    $\mathrm{KA}$ &$\vdash A^i_j \varphi\to \bigwedge_{k,l\in\mathcal{G}} C^k_l A^i_j \varphi$\\
    $\mathrm{NKA}$ &$\vdash \neg A^i_j \varphi\to \bigwedge_{k.l\in\mathcal{G}} C^k_l \neg A^i_j \varphi$\\
    $\mathrm{K_L}$ & 
    $\vdash L_j(\varphi\to \psi)\to 
    (L_j\varphi \to L_j\psi)$\\
    $\mathrm{T_L}$ & $\vdash L_j \varphi \to \varphi$\\
    $\mathrm{5_L}$ & $\vdash\neg L_j\varphi
    \to L_j \neg L_j\varphi$\\
    $\mathrm{K_{[\equiv]}}$ & 
    $\vdash [\equiv]^i_j(\varphi\to \psi)\to 
    ([\equiv]^i_j\varphi \to [\equiv]^i_j\psi)$\\
    $\mathrm{T_{[\equiv]}}$ & $\vdash [\equiv]^i_j \varphi \to \varphi$\\
    $\mathrm{5_{[\equiv]}}$ & $\vdash\neg [\equiv]^i_j\varphi
    \to [\equiv]^i_j \neg [\equiv]^i_j\varphi$\\
    $\mathrm{K_{C}}$ & $\vdash C^i_j(\varphi\to \psi)\to 
    (C^i_j\varphi \to C^i_j\psi)$\\
    MIX & $\vdash C^i_j\varphi \to \varphi
    \wedge[\equiv]^i_j L_j C^i_j\varphi$\\
    IND & 
    $\vdash C^i_j(\varphi\to [\equiv]^i_j L_j\varphi
    )\to
    (\varphi \to C^i_j\varphi)$\\
    KAC & $\vdash K^i_j\varphi\leftrightarrow A^i_j\varphi\wedge 
    C^i_j\varphi$\\\hline
    \multicolumn{2}{|c|}{Inference Rules}\\\hline
    MP & If $\vdash \varphi$ and $\vdash \varphi\to\psi$, then 
    $\vdash \psi$\\
    LG & If $\vdash \varphi$ then $\vdash L_j\varphi$\\
    $\mathrm{[\equiv] G}$ &
    If $\vdash \varphi$ then $\vdash [\equiv]^i_j\varphi$\\
    CG & If $\vdash \varphi$ 
    then $\vdash C^i_j\varphi$\\\hline
   \end{tabular}
 \end{table}

\subsection{SOUNDNESS}
\begin{thm}
If $\vdash\varphi$, then $\vDash\varphi$.
\end{thm}
\begin{proof}
  By induction on the construction of \textbf{ALP}, 
  we prove it for any formulas. 
  First, we prove that all axioms are valid. 
  For logical connectives, $L_j$, and $[\equiv]^i_j$
  can be proven similarly to those used in \textbf{S5}. 
  $A^i_j$ and $K^i_j$ are also straightforward.
  We show the proof of only $C^i_j$ here. 
  \begin{itemize}
  \item For $\mathrm{K_C}$, suppose that 
  $M,w\vDash  C^i_j(\varphi\to\psi)$, and 
  $M,w\vDash C^i_j\varphi$. 
  Since $M,v\vDash\varphi\to\psi$, and $M,v\vDash\varphi$ 
  for all $v$ such that $(w,v)\in (R_j \ \circ\equiv^i_j)^+$, 
  $M,v\vDash\psi$. Thus $M,w\vDash C^i_j\psi$.
  \item For MIX, suppose that $M,w\vDash  C^i_j\varphi$. 
  Since $(R_j \ \circ\equiv^i_j)^+$ is equivalent and the 
  transitive closure, $M,w\vDash\varphi\wedge
  [\equiv]^i_j L_j C^i_j\varphi$. 
  \item For IND, suppose that $M,w\vDash 
  C^i_j(\varphi\to [\equiv]^i_j L_j\varphi)$, 
  and $M,w\vDash\varphi$, then 
  for all $v$ such that $(w,v)\in (R_j \ \circ\equiv^i_j)^+$, 
  $M,v\vDash\varphi\to[\equiv]^i_j L_j\varphi$. 
  Thus $M,w\vDash[\equiv]^i_j L_j\varphi$. 
  It means $\varphi$ holds at all possible worlds 
  from $w$ on $R_j \ \circ\equiv^i_j$, and 
  $[\equiv]^i_j L_j\varphi$ holds even at that world. 
  Therefore, $M,w\vDash\varphi\to C^i_j\varphi$.
  \end{itemize}
  Then, it is enough to prove that 
  if the assumptions are valid,
  they are also valid for all inference rules.
  All of them are straightforward. 
\end{proof}

\subsection{COMPLETENESS}
In proof of the completeness theorem, 
we use the canonical model used in the proof 
on modal logic \cite{chellas1980modal}. 
However, in \textbf{ALP}, we can take a set of formulas, 
such as $\Phi = \{([\equiv]^i_j L_j)^n \varphi
\mid n\in \mathbb{N}\}
\cup \{\neg C^i_j\varphi\}$ for each $i,j\in\mathcal{G}$, 
where $([\equiv]^i_j L_j)^n$ is $n$ iterations of 
$[\equiv]^i_j L_j$. 
Therefore, our logic is no longer compact.
It is necessary to restrict canonical models 
to a finite set of formulas.  
This technique is used in proof on logic 
with common knowledge defined by 
the reflexive-transitive closure of relations. 
We customize the tools and techniques 
in \cite{van2007dynamic} for our logic and use them. 

First, we define \textit{closure} as a restricted set of formulas.
\begin{df}
  Let $cl : \mathcal{L}\to 2^{\mathcal{L}}$ 
  be the function such that for every $\varphi\in\mathcal{L}$ 
  and each $i,j\in\mathcal{G}$, 
  $cl(\varphi)$ is the smallest set satisfying that: 
  \begin{itemize}
    \item[1.] $\varphi\in cl(\varphi)$; 
    \item[2.] If $\psi\in cl(\varphi)$ then 
    $sub(\psi)\subseteq cl(\varphi)$
     where $sub(\psi)$ is the set of subformulas of $\psi$;
    \item[3.] If $\psi\in cl(\varphi)$ and 
    $\psi$ is not a form of negation, 
    then $\neg\psi\in cl(\varphi)$;
    \item[4.] If $A^i_j\psi\in cl(\varphi)$, 
    then $C^k_l A^i_j \psi$, $C^k_l \neg A^i_j \psi \in cl(\varphi)$;
    \item[5.] If $A^i_j\psi\in cl(\varphi)$, 
    then $[\equiv]^i_j p$,  $A^i_j \chi$, $A^k_l \psi \in cl(\varphi)$ 
    where $\chi\in sub(\psi)$ and $p$ is an atomic proposition in $cl(\varphi)$;
    \item[6.] If $L_j\psi\in sub(\varphi)$, 
    then $ L_j L_j\psi$ and 
    $L_j\neg L_j\psi\in cl(\varphi)$;
    \item[7.] If $[\equiv]^i_j\psi\in sub(\varphi)$, 
    then $[\equiv]^i_j[\equiv]^i_j\psi$ and 
    $[\equiv]^i_j\neg[\equiv]^i_j\psi\in cl(\varphi)$;
    \item[8.] If $C^i_j\psi\in cl(\varphi)$, 
    then $[\equiv]^i_j L_j C^i_j\psi
    \in cl(\varphi)$;
    \item[9.] If $L_k O\psi\in cl(\varphi)$, 
    then $L_k L_k O \psi$ and $L_k\neg L_k O\psi
    \in cl(\varphi)$ where $O\in\{A^i_j,C^i_j\}$;
    \item[10.] If $[\equiv]^i_jL_j C^i_j\psi\in cl(\varphi)$,
    then $ [\equiv]^i_j[\equiv]^i_j L_jC^i_j\psi$ and 
    $[\equiv]^i_j\neg[\equiv]^i_j L_jC^i_j\psi\in cl(\varphi)$;
    \item[11.] If $K^i_j\psi\in cl(\varphi)$, 
    then $A^i_j \psi$ and $C^i_j \psi\in cl(\varphi)$.
  \end{itemize} 
\end{df}
We call it the closure of $\varphi$. 

\begin{lemma}
  For every $\varphi$, $cl(\varphi)$ is finite.
\end{lemma}
\begin{proof}
  We prove it by induction on the structure of $\varphi$. 
  This proof is straightforward.   
\end{proof}

\begin{df}
  Let $\Phi$ be the closure of a formula. 
  $\Gamma$ is a maximal consistent set in $\Phi$ iff 
  \begin{itemize}
    \item[1.] $\Gamma \subseteq \Phi$; 
    \item[2.] $\Gamma \nvdash \bot$;
    \item[3.] There is no $\Gamma'$ such that 
    $\Gamma\subset\Gamma'$ 
    and $\Gamma'\nvdash \bot$;
  \end{itemize}
\end{df}

\begin{lemma}
  Let $\Phi$ be the closure of a formula. 
  If $\Gamma$ is a consistent set in $\Phi$, 
  then there exists a maximal consistent set $\Delta$ 
  in $\Phi$ 
  such that $\Gamma\subseteq\Delta$. 
\end{lemma}
\begin{proof}
  It follows immediately from the property that $\Phi$ is finite.
\end{proof}
Then, a maximal consistent set 
can be generated at any time from a consistent set. 

Next, the base model for a restricted set of formulas is defined as follows.  
\begin{df}
  Let $\Phi$ be the closure of a formula.
  The base model $C(\mathrm{ALP})$ 
  for $\Phi$ is a tuple $\langle C(W), \{C(R_j)\}_{j\in\mathcal{G}}, C(V),
  \{C(\equiv^i_j)\}_{i,j\in\mathcal{G}}\rangle$, where:
\begin{itemize}
  \item $C(W)\coloneqq\{\Gamma \mid \Gamma \text{ is 
  a maximal consistent set in }\Phi\}$; 
  \item $(w,v)\in C(R_j) \textit{\  iff  \ }
  \{\varphi\mid L_j\varphi\in w\}\subseteq v$;
  \item $C(V)(p)\coloneqq \{\Gamma \mid p\in\Gamma\}$;
  \item $(w,v) \in C(\equiv^i_j)\textit{\  iff  \ } 
  \{\varphi\mid [\equiv]^i_j\varphi\in w\}
  \subseteq v.$
\end{itemize}
\end{df}

\begin{df}
Let $\Phi$ be the closure of a formula and $\Lambda$ be a maximal consistent set in $\Phi$.
The divided model $C_{\Lambda}(\mathrm{ALP})$ by $\Lambda$ for $\Phi$ is a tuple $\langle 
C_{\Lambda}(W), \{C_{\Lambda}(R_j)\}_{j\in\mathcal{G}}, C_{\Lambda}(V),
\{C_{\Lambda}(\mathscr{A}^i_j)\}_{i,j\in\mathcal{G}},$
$\{C_{\Lambda}(\equiv^i_j)\}_{i,j\in\mathcal{G}}\rangle$, where:
  \begin{itemize}
    \item $C_{\Lambda}(W)\coloneqq\{\Gamma \mid \Gamma \text{ is 
    a maximal consistent set in }\Phi$, and 
    $(\Lambda, \Gamma)\in \bigcup_{i,j\in \mathcal{G}} 
    (C(R_j)\circ C(\equiv^i_j))^+\}$; 
    \item $C_{\Lambda}(R_j) \coloneqq C(R_j) \cap (C_{\Lambda}(W)\times C_{\Lambda}(W))$;
    \item $C_{\Lambda}(V)(p)\coloneqq C(V)(p)\cap C_{\Lambda}(W)$;
    \item $C_{\Lambda}(\mathscr{A}^i_j)\coloneqq \{p\mid 
    \text{for all } w\in C_{\Lambda}(W), A^i_j p\in w\}$;
    \item $C_{\Lambda}(\equiv^i_j)\coloneqq C(\equiv^i_j)\cap (C_{\Lambda}(W)\times C_{\Lambda}(W))$.
  \end{itemize}
\end{df}

\begin{lemma}
For every $\varphi$, each divided model by $\Lambda$
for the closure of $\varphi$ 
is an epistemic model with awareness, where $\Lambda$ is a maximal consistent set in the closure.
\end{lemma}
\begin{proof}
We prove that each divided model by $\Lambda$
for the closure of $\varphi$ 
satisfies 
the definition of an epistemic model with awareness. 
\begin{itemize}
\item For $C(R_i)$, it can be proven in 
the same proof strategy as \textbf{S5}.
\item For $C_{\Lambda}(\mathscr{A}^i_j)$, it is enough 
to prove that if $p\in C_{\Lambda}(\mathscr{A}^i_j)$ then 
$p\in C_{\Lambda}(\mathscr{A}^i_i)$ for every $p\in\mathcal{P}$.
Suppose $p\in C_{\Lambda}(\mathscr{A}^i_j)$, then 
for all $w\in C_{\Lambda}(W)$, $A^i_j p\in w$. 
Thus, it follows that $A^i_i p\in w$ for all $w\in C_{\Lambda}(W)$ from $\mathrm{AI}$.
\item For $C_{\Lambda}(\equiv^i_j)$, it is enough to prove 
that for all $(w,v)\in C_{\Lambda}(\equiv^i_j)$, 
for every $p\in C_{\Lambda}(\mathscr{A}^i_j)$, 
if $w\in C_{\Lambda}(V)(p)$, then $v\in C_{\Lambda}(V)(p)$, and vice versa.
From left to right, suppose that $w\in C_{\Lambda}(V)(p)$ 
for every $p\in C_{\Lambda}(\mathscr{A}^i_j)$. 
Then, $p\in w$ and $A^i_j p\in w$. 
$p\in v$ follows from $\mathrm{AN[\equiv]}$. 
The reverse direction is 
proven by $C_{\Lambda}(\equiv^i_j)$ is equivalent
.
\end{itemize} 
\vspace{-5mm}
\end{proof}

We introduce $C^i_j$-paths. 
\begin{df}
Let $\Phi$ be the closure of a formula.
A $C^i_j$-path from $\Gamma$ is a 
sequence $\Gamma_0,\cdots,\Gamma_n$ of maximal
consistent sets in $\Phi$ such that 
$(\Gamma_k,\Gamma_{k+1})
\in  C(R_j) \ \circ C(\equiv^i_j)$ for all $k$, 
where $0\leq k\leq n$, and $\Gamma_0 = \Gamma$. 
The length of $\Gamma_0,\cdots,\Gamma_n$ is $n$.
A $\varphi$-path is a sequence 
$\Gamma_0,\cdots,\Gamma_n$ of maximal
consistent sets in $\Phi$ 
such that $\varphi\in\Gamma_k$ for all $k$,  
where $0\leq k\leq n$.
\end{df}  

\begin{lemma}
Let $\Phi$ be the closure of a formula and
 $\Gamma, \Delta$ be a maximal consistent set in $\Phi$.
If $\bigwedge\Gamma \wedge \neg[\equiv]^i_j L_j 
\neg \bigwedge\Delta$ is consistent, then $(\Gamma,\Delta)\in (C(R_j) \ \circ C(\equiv^i_j))$. 
\end{lemma}

\begin{proof}
Suppose that $\bigwedge\Gamma \wedge \neg[\equiv]^i_j L_j \neg \bigwedge\Delta$ 
is consistent, and $[\equiv]^i_j L_j\varphi \in \Gamma$ for every $\varphi$. 
Then, $[\equiv]^i_j L_j\varphi \wedge \neg[\equiv]^i_j 
L_j \neg \bigwedge\Delta$ is 
consistent. 
If $\varphi\not\in\Delta$, $\neg\varphi\in\Delta$. 
It follows that 
$[\equiv]^i_j L_j\varphi \wedge\neg[\equiv]^i_j 
L_j\varphi$ is consistent, but this formula is a contradiction. 
Thus, $\varphi\in\Delta$. 
\end{proof}

\begin{lemma}
Let $\Phi$ be the closure of a formula and 
$\Gamma, \Delta$ be maximal consistent sets in $\Phi$.
If $C^i_j\varphi\in\Phi$, then $C^i_j\varphi\in\Gamma$ iff 
every $C^i_j$-path from $\Gamma$ is a $\varphi$-path and a $C^i_j \varphi$-path. 
\end{lemma}

\begin{proof}
($\Rightarrow$) We prove it by induction on the length of a $C^i_j$-path.
\begin{itemize}
\setlength{\itemsep}{-1pt}
\item For the base case, suppose that the length 
of a $C^i_j$-path is $0$, 
$C^i_j \varphi\in\Phi$, and $C^i_j\varphi\in\Gamma$. 
Then $\Gamma =\Gamma_0 = \Gamma_n$. 
By MIX, $\varphi\in\Gamma$.
\item For induction steps, 
suppose that the length of a $C^i_j$-path is $k+1$, 
$C^i_j \varphi\in\Phi$ and $C^i_j\varphi\in\Gamma$.
By the induction hypothesis, $C^i_j\varphi\in\Gamma_k$.
Since MIX and the definition of $C(R_j)$ and $C(\equiv^i_j)$, 
$\varphi$ and $C^i_j\varphi\in \Gamma_{k+1}$.
\end{itemize}

($\Leftarrow$) Let $S(C^i_j,\varphi)$ be a set of 
maximal consistent sets $\Delta$ in $\Phi$ such that 
every $C^i_j$-path from $\Delta$ is a $\varphi$-path. 
We introduce a special formula: 
\[\chi = \bigvee_{\Delta\in S(C^i_j,\varphi)}\bigwedge \Delta.\] 

Suppose that every $C^i_j$-path from $\Gamma$ is a $\varphi$-path. 
First, we need to prove these three: 
\begin{align*}
(1)\ \vdash &\bigwedge\Gamma\to \chi; 
\quad (2)\ \vdash \chi\to\varphi;
\quad (3)\ \vdash \chi\to [\equiv]^i_j L_j\chi.
\end{align*}

\begin{itemize}
\setlength{\itemsep}{-1pt}
\item For (1), $\Gamma\in S(C^i_j,\varphi)$ by the assumption. 
Thus, $\vdash \bigwedge\Gamma\to \chi$. 
\item For (2), since every 
$C^i_j$-path from $\Delta$ is a $\varphi$-path, 
$\varphi\in\Delta$ for every $\Delta \in S(C^i_j,\varphi)$. 
Thus, $\varphi$ is derived from $\chi$. 
\item For (3), we prove it by contradiction.
Suppose $\chi\wedge\neg[\equiv]^i_j L_j\chi$ is consistent. 
By the construction of $\chi$, there exists $\Delta$ such that 
$\bigwedge\Delta\wedge\neg[\equiv]^i_j L_j\chi$ is consistent. 
The set $\neg\bigvee_{\Theta\in C(W)\setminus S(C^i_j,\varphi)}
\bigwedge\Theta$ is equivalent to $\chi$
because the disjunction of the complement of the other combinations 
can express a particular set of combinations represented by $\chi$. 
Therefore, $\bigwedge\Delta\wedge\neg[\equiv]^i_j L_j
\neg\bigvee_{\Theta\in C(W)\setminus S(C^i_j,\varphi)}
\bigwedge\Theta$ is consistent. There is $\Theta$ such that 
$\bigwedge\Delta\wedge\neg[\equiv]^i_j L_j
\neg\bigwedge\Theta$ is consistent. 
By Lemma 4, $(\Delta,\Theta)\in  C(R_j) \ \circ C(\equiv^i_j)$. 
There exists a $C^i_j$-path from $\Delta$ 
that is not a $\varphi$-path. 
This is a contradiction. 
Thus, $\vdash \chi\to [\equiv]^i_j L_j\chi$. 
\end{itemize}

By $(3)$ and CG, $\vdash C^i_j(\chi\to [\equiv]^i_j L_j\chi)$.
It follows that $\vdash\chi\to C^i_j \chi$ from IND. 
By $(1)$ and $(2)$, $\vdash\bigwedge \Gamma\to C^i_j\varphi$.
Thus, $C^i_j\varphi\in\Gamma$.    
\end{proof}

\begin{lemma}
Let $\Phi$ be the closure of a formula and 
$C_{\Lambda}(\mathrm{ALP})$ be a divided model for $\Phi$. 
For all $w\in C_{\Lambda}(W)$ and every $\varphi\in \Phi$, 
$C_{\Lambda}(\mathrm{ALP}),w\vDash\varphi$ iff $\varphi\in w$.
\end{lemma}
\begin{proof}
We prove it by induction on the structure of formulas. 
The cases other than $A^i_j$ and $C^i_j$ 
are trivial, including the base case. For $L_j \varphi$ and $[\equiv]^i_j$, it is proven in the same proof strategy as \textbf{S5}.
\begin{itemize}
\item For the case of $A^i_j \varphi$, 
we prove it by induction on the structure of $\varphi$.
Suppose that $C_{\Lambda}(\mathrm{ALP}),w\vDash A^i_j\varphi$, 
then $At(\varphi)\subseteq C_{\Lambda}(\mathscr{A}^i_j)$. 
It means for every $p\in\mathcal{P}$ and for all $v\in C_{\Lambda}(W)$, 
if $p\in At(\varphi)$ then $A^i_j p \in v$. 
\begin{itemize}
\item For the base case, 
$A^i_j p \in v$ for all $v\in C_{\Lambda}(W)$, since $p \in At(p)$.
 Thus, $A^i_j p \in w$. 
  \item For the other cases, 
  we obtain the desired proof by induction hypothesis 
  and decomposing the formula with corresponding axioms: 
  AN, AC, AA, $\mathrm{A[\equiv]}$, ACM, AL, and AK.
\end{itemize}
The reverse direction is proven by $\mathrm{KA}$, $\mathrm{NKA}$, and $\mathrm{MIX}$. 
  \item For the case of $C^i_j$,   
  suppose that $C_{\Lambda}({\mathrm{ALP}}),w\vDash C^i_j\varphi$. 
  Then, $C_{\Lambda}({\mathrm{ALP}}),v\vDash \varphi$ for all $v$ 
  such that $(w,v)\in ( C_{\Lambda}(R_j) \ \circ C_{\Lambda}(\equiv^i_j))^+$. 
  It means every $C^i_j$-path from $w$ is a $\varphi$ and $C^i_j\varphi$-path.
  By Lemma 5, $C^i_j\varphi\in w$. 
  The reverse direction is proven similarly.
\end{itemize}
\vspace{-5mm}
\end{proof}

\begin{lemma}
Let $\Phi$ be the closure of a formula and 
$\Gamma$ be a maximal consistent set in $\Phi$.
For every $\varphi\in\Phi$ and
every maximal consistent set $\Gamma$, 
if $\varphi\in \Gamma$ then $\vdash\varphi$.
\end{lemma}
\begin{proof}
We prove it by contraposition. Suppose $\nvdash\varphi$. By Lemma 2, there is a maximal consistent set in $\Phi$ that does not contain $\varphi$. Thus, $\varphi\not\in\Gamma$.
\end{proof}

\begin{thm}
For every $\varphi\in\mathcal{L}_{\mathcal{P}}$, 
if $\vDash\varphi$, then $\vdash\varphi$.
\end{thm}
\begin{proof}
  Suppose that $\vDash\varphi$, then $C_{\Lambda}(\mathrm{ALP}),w\vDash\varphi$ for every divided model $C_{\Lambda}(\mathrm{ALP})$ for the closure of $\varphi$ by Lemma 3. $\varphi\in w$ by Lemma 6. Thus, $\vdash\varphi$ by Lemma 7.
\end{proof}

\section{EPISTEMIC ACTIONS}
In epistemic logic, including logic of awareness, 
we formalize how the information held by agents changes 
for applications and understanding of concepts. 
In this paper, we introduce two actions that are 
`becoming aware of' and `becoming unaware of' 
as preparation for incorporating agent communication.
These are the basic actions relevant to `awareness of'
\cite{van2010dynamics}.

First, we add two new operators to syntax, which 
are $[+\varphi]^i_j$ and $[-\varphi]^i_j$ for each $i,j\in\mathcal{G}$. 
$[+\varphi]^i_j\psi$ reads `$j$ 
become aware of $\varphi$ in $i$'s viewpoint'.
$[-\varphi]^i_j\psi$ reads `
$j$ become unaware of $\varphi$ in $i$'s 
viewpoint.'
We extend the satisfaction relation of $\mathcal{ALP}$ 
as follows:
\begin{align*}
	M,w &\vDash [+\varphi]^i_j\varphi
	  \textit{\  iff  \ } M[+\varphi]^i_j,w\vDash \varphi;\\[-2pt]
	M,w &\vDash [-\varphi]^i_j\varphi
    \textit{\  iff  \ } M[-\varphi]^i_j,w\vDash \varphi. 
\end{align*} 
Formulas with dynamic operators 
are evaluated in the updated models, 
which are $M[+\varphi]^i_j$ and $M[-\varphi]^i_j$. 
We define these as follows: 
\begin{df}
$M[+\varphi]^i_j$ is a tuple 
$\langle W, \{R_k\}_{k\in\mathcal{G}}, V,$
$\{\mathscr{A}[+\varphi]^k_l\}_{k,l\in\mathcal{G}},$ 
$\{\equiv^k_l\}_{k,l\in\mathcal{G}}\rangle$, where: 
\begin{align*}
  \ &\mathscr{A}[+\varphi]^k_l \coloneqq 
  \begin{cases}
   \mathscr{A}^k_l\cup \{At(\varphi)\} & k=i \text{ \ and \ }
    l=j,\\
   \mathscr{A}^k_l & \text{otherwise}.
  \end{cases}
\end{align*}
$M[-\varphi]^i_j$ is a tuple 
$\langle W, \{R_k\}_{k\in\mathcal{G}}, V,
\{\mathscr{A}[-\varphi]^k_l\}_{k,l\in\mathcal{G}}$,
  $\{\equiv^k_l\}_{k,l\in\mathcal{G}}\rangle$, where: 
\begin{align*}
	\ &\mathscr{A}[-\varphi]^k_l \coloneqq 
	  \begin{cases}
	   \mathscr{A}^k_l\setminus \{At(\varphi)\} & k=i \text{ and } l=j,\\
	   \mathscr{A}^k_l & \text{otherwise}.
	  \end{cases}
\end{align*}
\end{df}
For example, $[+n]^b_b K^b_b K^b_a p_b$ is true at $w_1$ 
in the example in Section 3. 

In order to provide the corresponding Hilbert-system, 
there is a technique to prove the completeness theorem of 
logic with dynamic operators, such as PAL 
(Public Announcement Logic) \cite{plaza1989logics}.
The technique replaces a formula with a dynamic operator 
of a simple formula that is logically equivalent. 
For example, 
$[+\varphi]L_j\psi\leftrightarrow 
L_j[+\varphi]\psi$ holds for $L_j$ operator. 
For $[\equiv]^i_j$ and $C^i_j$ operators, 
we refer to \cite{grossi2015ceteris} for 
identifying possible worlds with the same valuation 
for atomic propositions in the changed awareness set. 
The logic in \cite{grossi2015ceteris} 
has the operator that represents a proposition holds at 
the possible world whose the same valuation for 
all the elements of a particular set of atomic propositions. 
We leave this part as our future work. 
  
\section{RELATED WORK}
We introduce some logic or ideas relevant to 
our logic. 
\cite{van2009awareness,ditmarsch2011becoming} 
is based on a similar idea as this paper, which is 
to connect agents' state of awareness 
with the distinctions of possible worlds. 
The main difference is that our logic can represent 
not only the distinctions of possible worlds 
but also possible worlds searched according to the distinction. 
Unlike our logic, these logics can search even for worlds with 
a different valuation for propositions of 
which an agent is unaware 
but do not adjust accessibility. 

\textit{Team semantics} used in dependence 
logic \cite{sano2015axiomatizing} also has 
a similar idea that formulas are 
true in a specific group of possible worlds. 
This semantics has a structure that a subset 
of possible worlds called a \textit{term} supports a formula.

As for epistemic actions for awareness, 
several papers are using the idea of 
PAL \cite{plaza1989logics}, including this paper. 
In particular, \cite{grossi2015syntactic,fernandez2021awareness} 
proposed a realistic formalization of epistemic actions, 
such as updating awareness by inference. 
The idea of an \textit{action model} \cite{baltag1998logic} 
might help formalize agent communication in our logic. 
Action models control communicative actions separately 
from a Kripke model that decides knowledge. 
It allows us to formalize complex actions, 
such as a misleading private announcement.

Semantic approaches to awareness are active 
in the field of economics. 
It is also called the event-based approach, in which the concept 
of \textit{events} that are a set of possible worlds is introduced, 
and knowledge is expressed as an operator on events. 
The logic system proposed in \cite{modica1994awareness} is 
the early work of the approach. 
\cite{halpern2001alternative} found it to be equivalent to 
a part of the logic in \cite{fagin1988belief}. 
Since the work of \cite{modica1999unawareness}, 
the focus has been on a formalization of the concept of unawareness 
\cite{heifetz2006interactive,heifetza2008canonical}. 

\section{CONCLUSION}
In this paper, we have introduced 
Awareness Logic with Partition ($\mathcal{ALP}$), 
where we incorporated the notion of partition among possible worlds 
and have extended the distinction of aware/unaware propositions 
to indistinguishable possible worlds. 
With this, we have properly reflected the agent's 
awareness to other agents' knowledge. 
Employing this framework, we have shown an example 
where the behavior of each agent could be logically explained. 

Our contributions of this logic are two-fold. 
From the logical viewpoint, 
we introduced the syntax and the semantics of $\mathcal{ALP}$
and have shown its completeness. 
From the viewpoint of applicability to real world, 
we have shown the architecture to explain 
the strategic behavior of rational agents in a society 
or game theory.
We expect that the logic offers a foundation 
for formal expressions of human minds and practical agent communication. 

There are several directions in the future. 
On the conceptual side, 
we consider incorporating more epistemic actions and concepts, 
such as common knowledge, to represent practical agent communication. 
On the technical side, the axiomatic system 
of the dynamic extension, discussed in Section 5, remains.
In addition, our logic is applicable to the studies dealing 
with multiple agents' reasoning, such as
description and analysis of games that take into account 
players' awareness of possible strategies 
\cite{feinberg2005games,kaneko2002bounded}. 
Specifically, we plan to use the logic to analyze rationality 
to reach an equilibrium in games with awareness. 

\section*{ACKNOWLEDGMENTS}
The authors thank Professor Thomas \AA gotnes and the anonymous reviewers 
for their many insightful comments. 
This work was supported by JSPS kaken 22H00597.

\bibliographystyle{apalike}
{\small
\bibliography{refs}}

\end{document}